\documentclass[12pt,journal,onecolumn]{IEEEtran}

\usepackage{amsmath}
\usepackage{amssymb,cool}
\usepackage{graphicx,cite}

\newtheorem{prop}{\textbf{Proposition}}
\newtheorem{lemma}{\textbf{Lemma}}
\newtheorem{theorem}{\textbf{Theorem}}

\newtheorem{remark}{Remark}

\newcommand{\sinr}{{\rm SINR}}
\newcommand{\snr}{{\rm SNR}}

\newcommand{\pout}{{\mathcal{F}(\beta, \alpha)}}
\newcommand{\pkout}{{\mathcal{F}^{(k)}(\beta_k, \alpha)}}

\IEEEoverridecommandlockouts

\begin{document}
\title{Downlink SDMA with Limited Feedback in Interference-Limited Wireless Networks}

\author{\authorblockN{Marios Kountouris and Jeffrey G. Andrews} \\
\thanks{M. Kountouris is with SUPELEC, France, Email: \texttt{marios.kountouris@supelec.fr}, and J. G. Andrews is with the University of Texas at Austin, USA, Email: \texttt{jandrews@ece.utexas.edu}. This research has been supported by the DARPA IT-MANET program.
}
}

\maketitle

\begin{abstract}
The tremendous capacity gains promised by space division multiple access (SDMA) depend critically on the accuracy of the transmit channel state information. In the broadcast channel, even without any network interference, it is known that such gains collapse due to interstream interference if the feedback is delayed or low rate.  In this paper, we investigate SDMA in the presence of interference from many other simultaneously active transmitters distributed randomly over the network.  In particular we consider zero-forcing beamforming in a decentralized (ad hoc) network where each receiver provides feedback to its respective transmitter. We derive closed-form expressions for the outage probability, network throughput, transmission capacity, and average achievable rate and go on to quantify the degradation in network performance due to residual self-interference as a function of key system parameters.  One particular finding is that as in the classical broadcast channel, the per-user feedback rate must increase linearly with the number of transmit antennas and SINR (in dB) for the full multiplexing gains to be preserved with limited feedback. We derive the throughput-maximizing number of streams, establishing that single-stream transmission is optimal in most practically relevant settings.  In short, SDMA does not appear to be a prudent design choice for interference-limited wireless networks. 
\end{abstract}

\section{\label{sec:intro} Introduction}

In multiuser MIMO (multiple-input, multiple-output) channels, the spatial multiplexing capability offered by multiple antennas can be advantageously exploited to significantly increase the achievable throughput. In single-cell point-to-multipoint channels, the achievable sum rate scales linearly with the number of transmit antennas, even when the mobile users have only a single antenna. 
By duality, linear increase with the number of receiver antennas can also be achieved in multipoint-to-point channels, even with a single-antenna transmitter.
Extensive research on MIMO broadcast channels over the last few years has revealed that the capacity can be boosted by transmitting to multiple users simultaneously, by means of Space Division Multiple Access (SDMA), rather than trying to maximize the capacity of a single-user link \cite{CaiSha03, ShiftingMIMO}. 
Nevertheless, all these promising gains critically depend on accurate channel state information (CSI), and in contrast to point-to-point channels, the quality of CSI affects the multiplexing gain of multiuser MIMO systems. 
As a result, a considerable amount of effort has been dedicated to multiuser MIMO systems operating with partial CSI at the transmitter (CSIT) in the absence of out-of-cell interference \cite{LFB_JSAC_Tutorial}. 

In this paper we are interested in the capacity gains that SDMA may provide in decentralized (ad hoc) networks with both intra-cell due to imperfect CSIT and other user interference due to uncoordinated concurrent transmissions. We aim at answering whether and how aggressive use of multiple antennas through SDMA may increase the network throughput under a broad set of scenarios.  
Specifically, we build upon the practically relevant limited feedback model, in which each user is allowed to feed back $B$-bit quantized information on its channel direction based on a predetermined codebook known at both the transmitter and the receivers.
Due to the high implementation complexity and sensitivity to channel errors of the optimal scheme (dirty paper coding \cite{Cos83}), linear precoding based on zero-forcing beamforming (ZFBF) \cite{SpeSwi04} is employed here. ZFBF has been shown to achieve full multiplexing gain while exhibiting reduced complexity \cite{Dimic, Yoo05}. ZFBF is also used both for its asymptotic optimality in MIMO point-to-multipoint channels and its analytical tractability. In short, it is a logical starting point for understanding the effect of imperfect CSIT in interference-limited networks.

\subsection{Related Work}
Several recent papers have studied MIMO ad hoc networks with Poisson distributed interferers, but the majority of prior work considers only point-to-point ad hoc links, in which each transmitter communicates with only one receiver at a time (TDMA). 
Several receive antenna processing techniques have been investigated quantifying the performance gains of antenna combining and interference cancelation~\cite{HunAnd08, HuaAndSub, PZF, Olfa_MMSE10}. 
Open-loop spatial multiplexing with linear receivers is studied in~\cite{Louie09}, while \cite{Vaze09} studies multi-mode precoding with receive interference cancelation. Spatial multiplexing with limited transmit CSIT have been studied for asymptotically large number of antennas in \cite{Gov10limCSI}. 
Multiuser transmission, in which each transmitter sends different messages to multiple users, is analyzed recently for scalar broadcast channels using superposition coding in \cite{MartinISIT10}. The performance of multiuser MIMO communication in a Poisson field of interferers is first studied in \cite{myITW09} considering non-linear and linear precoding with perfect CSIT. Therein the network-wide capacity is shown to increase linearly and sublinearly with the number of antennas when DPC and ZFBF are employed, respectively, for single-antenna receivers. It is also shown that it is often throughput maximizing to send fewer streams than the number of transmit antennas. Here, we investigate whether these results derived under the idealized perfect CSI assumption still hold in a more practical and realistic scenario. 
Single-user beamforming is compared to SDMA with perfect CSIT in \cite{VikramTWC09} for two-tier networks with spatial randomness. A key finding is that single-stream transmission at each tier provides significantly superior coverage and spatial reuse relative to multiuser transmission.

\subsection{Main Contributions}
In this paper, in contrast to all prior work on multiuser MIMO ad hoc networks, we consider SDMA ad hoc links in which only local, partial CSI is known at the transmitter.  We investigate the zero-forcing transmission technique with quantized CSIT in the context of decentralized interference-limited networks using a random access medium access control protocol. 
The spatial distribution of the nodes follows a homogeneous Poisson point process (PPP), and each transmitter serves multiple receivers (point-to-multipoint), each located at a certain distance away from it. 

First, we investigate the performance of limited feedback SDMA and derive novel closed-form expressions for the outage probability, network aggregate throughput, transmission capacity \cite{WebYan05}, and mean user rate using stochastic geometric tools. These expressions enable us to quantify the capacity gains that zero-forcing precoding can provide in decentralized networks and the network-wide performance loss due to residual (uncanceled) multiuser interference caused by quantized channel information. Key findings are that there is a node density value that maximizes the network throughput and that for practically relevant values of feedback bits, if the number of antennas/streams is increased, the outage constraint for non zero node density is prohibitively high.

Second, we evaluate the performance degradation as a function of the feedback rate, the number of antennas, and the outage constraints for fixed feedback levels. A first result is that, similar to the single-cell case, the throughput of feedback-based zero-forcing is self-interference-limited, i.e. the average throughput is bounded if the feedback load is kept fixed, even if the transmit power is taken to infinity. 
We then provide the scaling of feedback bits in order to guarantee bounded performance offset. A key finding is that per-user feedback load $B$ must be increased almost linearly with the number of antennas and logarithmically with the target signal-to-interference-plus-noise ratio (SINR), i.e. similar to the single-cell case analyzed by Jindal \cite{Jindal_finiteFB} if the transmit power is replaced with the target SINR constraint.

Finally, we derive the optimal number of streams in order to maximize throughput and transmission capacity, enabling us to propose transmission schemes that dynamically adjust the number of streams to the network operating values. Our results establish that the optimality of single-stream transmission in most practically relevant settings. 
The main takeaway of this paper is that limited feedback SDMA may not be a wise use of an antenna array in ad hoc networks, and in general in network settings where interferers can be close by as in heterogeneous networks including femtocells, picocells, relays, and WiFi hotspots.



\section{System Model and Preliminaries}
\label{sec:model}

The network model consists of transmitters arranged according to a homogeneous Poisson point process (PPP) $\Phi$ of intensity $\lambda$ in $\mathbb{R}^2$. Each transmitter has $M$ antennas and communicates with a set of intended single-antenna receivers $\mathcal{K}$ with cardinality $\left|\mathcal{K}\right|=K \leq M$. Every transmitter sends $K$ streams destined to a different user each, forming thus a $K$-user broadcast cluster. Users are distributed according to some independent stationary point process. Transmissions are uncoordinated (random access MAC protocol) and the signal is attenuated according to the standard power law, i.e. the received power decays with the distance $d$ as $d^{-\alpha}$ for path loss exponent $\alpha > 2$. For the fading model (random channel component), we assume that all point-to-point channels experience i.i.d. Rayleigh block fading with unit mean. 

We investigate multiuser transmission, in which each transmitter sends a different message to its associated receivers. Due to the stationarity and the rotational invariance of the PPP (Slivnyak's Theorem~\cite{ StoKen96}), it is sufficient to analyze the performance of a typical one-to-many communication channel, as the statistics of the signal reception remain the same for the disjoint set of intended receivers in the broadcast cluster. Denote the typical transmitter by $T_0$ located at the origin communicating with its $k$-th typical user, located at distance $d_k$, for $k \in \mathcal{K}$ and denoted as $R_0^{(k)}$.

The received signal $y_{0k}$ at typical receiver $R_0^{(k)}, k \in \mathcal{K}$ assuming frequency-flat channels is given by
\begin{equation}
y_{0k} = \sqrt{\rho} \ d_{k}^{-\frac{\alpha}{2}}\mathbf{h}_{0k}\mathbf{x}_{0} + \sqrt{\rho}\sum_{i \in \Phi(\lambda)/T_0}D_i^{-\frac{\alpha}{2}}\mathbf{h}_{ik}\mathbf{x}_i + n_{0k}
\end{equation}
where $\rho = \frac{P}{K}$, $P$ is the transmit power, $D_i \in \mathbb{R}^2$ is the distance to the $i$-th transmitter, and $n_{0k} $ is the complex additive Gaussian noise with variance $\sigma^2$.
The vector channel from the $i$-th transmitter to $R_0^{(k)}$ is denoted by $\mathbf{h}_{ik} \in \mathbb{C}^{1\times M}$ and $\mathbf{x}_i$ is the $M \times 1$ normalized transmit signal vector of the $i$-th transmitter.
The vectors $\mathbf{h}$ are assumed to have i.i.d. $\mathcal{CN}(0,1)$ entries, independent across transmitters and of the random distances $D_i$. The index `0' is dropped for notation simplification, as all the subsequent analysis is performed on a typical broadcast cluster.

\subsection{Finite Rate Feedback Model}
We assume that each receiver has perfect knowledge of the channel to its corresponding transmitter. In each cluster, the transmitter acquires partial CSIT that only captures the spatial direction information of the channel, referred to as channel direction information. For that, a quantization codebook $\mathcal{V}_k=\{\mathbf{v}_{k1}, \mathbf{v}_{k2}, \ldots, \mathbf{v}_{kN}\}$ containing $N=2^{B}$ unit norm vectors $\{\mathbf{v}_{ki}\}_{i=1}^{N} \in \mathbb{C}^M$ is employed, assumed to be known to both $T_0$ and receiver $R_0^{(k)}$. At each feedback reporting slot, each receiver $k$ quantizes its channel realization $\mathbf{h}_k$ to the closest codeword with respect to the chordal distance \cite{Love_Grass,Mukkavilli},
\begin{equation*}
\hat{\mathbf{h}}_k = \arg\max_{\mathbf{v}_{ki}\in \mathcal{V}_k} |\bar{\mathbf{h}}_k \mathbf{v}_{ki}|^2 = \arg\max_{\mathbf{v}_{ki}\in \mathcal{V}_k} \cos^2(\angle(\bar{\mathbf{h}}_k,\mathbf{v}_{ki})),
\label{chordal_dist}
\end{equation*}
where $\bar{\mathbf{h}}_k = \mathbf{h}_k / \left\|\mathbf{h}_k\right\|$ corresponds to the channel direction. Each user sends the corresponding quantization index back to the transmitter using $B=\left\lceil \log_{2}N\right\rceil$ bits through an error and delay-free feedback channel\footnote{The error-free assumption can be well approximated using sufficiently powerful error-correcting codes over the feedback link, whereas the zero-delay assumption may be valid when the processing and feedback delays are small relative to the channel coherence time.}.
The optimal vector quantization strategy in multiuser downlink channels is not known in general, even in single-cell systems, and is out of the scope of our work. We resort hence to a vector quantization scheme following the quantization cell approximation (QCA) \cite{Mukkavilli, GiannakisVQ}.
It has been shown in \cite{GiannakisVQ, Yoo_JSAC} that QCA can facilitate the analysis and provide a very accurate performance approximation, with only small difference from random vector quantization.

\subsection{Zero-forcing Beamforming}
In this paper, we focus on linear precoding (downlink beamforming), in which the transmit symbol vector $\mathbf{x}$ is a linear function $\mathbf{x} = \sum_{k \in \mathcal{K}} \mathbf{w}_{k}s_{k}$, where $s_{k}$ is the data symbol intended for the $k$-th receiver and $\mathbf{w}_{k} \in \mathbb{C}^{M\times 1}$ is the unit-norm beamforming vector for user $k$. Specifically, users are served via zero-forcing for which the beamforming vectors are chosen as $\hat{\mathbf{h}}_{k}\mathbf{w}_{i} = 0$, $\forall k \neq i, k\in \mathcal{K}$. 

Let $\mathbf{H}(\mathcal{K}) = \left[\hat{\mathbf{h}}_1^{\rm T}, \ldots, \hat{\mathbf{h}}_K^{\rm T}\right]^{\rm T}$ denote the concatenation of the quantized channel vectors upon which zero-forcing is performed. The beamforming vectors are given by the Moore-Penrose pseudoinverse 
\begin{equation}
\mathbf{W}(\mathcal{K}) = \mathbf{H}(\mathcal{K})^{\dagger} = \mathbf{H}(\mathcal{K})^{\rm H}(\mathbf{H}(\mathcal{K})\mathbf{H}(\mathcal{K})^{\rm H})^{-1}
\end{equation}
with $\mathbf{w}_{k}$ obtained by normalizing the $k$-th column of $\mathbf{W}(\mathcal{K})$.
The receive SINR at the $k$-th typical user, treating interference as noise and using equal power allocation for each of the data streams, can be expressed as
\begin{equation*}
\sinr_k = \frac{\rho \left|\mathbf{h}_{k}\mathbf{w}_{k}\right|^2 d_k^{-\alpha}}{I_p  + I_q + \sigma^2},
\label{eq:SINRdef}
\end{equation*}
\begin{minipage}{0.43\linewidth}
\begin{eqnarray}
\text{with} \hspace{8mm} I_p = \displaystyle \sum_{i \in \Phi(\lambda)/T_0}\rho \left\|\mathbf{h}_{ik}\mathbf{W}_{i}\right\|^2 D_i^{-\alpha} \label{eq:Ip} 
\end{eqnarray}
\end{minipage}
\ \ \ \text{and}
\begin{minipage}{0.43\linewidth}
\begin{eqnarray}
I_q = \displaystyle \sum_{j \in \mathcal{K}, j \neq k} \rho \left|\mathbf{h}_{k}\mathbf{w}_{j}\right|^2d_k^{-\alpha} \label{eq:Iq}
\end{eqnarray}
\end{minipage} \vspace{1mm}\\
where $I_p$ is the aggregate inter-cluster interference from the Poisson field of interferers $\Phi/T_0$ and $I_q$ is the intra-cluster self-interference due to the fact that zero-forcing vectors are calculated based on quantized CSIT.

\subsection{Performance Metrics}
\emph{Outage probability.} A primary performance measure is the outage probability, which is defined as the probability that the received SINR falls below a target SINR $\beta$, i.e.
\begin{equation}
\pout = \mathbb{P}\left(\sinr \leq \beta \right).
\label{pout}
\end{equation}
It can be thought of equivalently as the probability of no coverage of a user, and is evidently a continuous increasing function of the intensity $\lambda$.\footnote{In an SDMA setting, we can alternatively define $\pout = \mathbb{P}\left(\mathcal{I}(\mathbf{x}_0;\mathbf{y}_0) \leq r \right)$, where $\mathcal{I}(\mathbf{x}_0;\mathbf{y}_0)$ is the mutual information between $\mathbf{x}_0$ and $\mathbf{y}_0$, and $r$ is a certain target information rate. However, a decomposed per-stream/user outage constraint is more meaningful in SDMA ad hoc networks, in which each stream contains a different message.}
In multiuser communication, different SINR statistics may be seen on different users (streams), resulting in a per-user outage probability $\pkout$, $\forall k \in \mathcal{K}$, with $\beta_k$ being the target SINR on stream $k$. 


\emph{Network throughput.} The network throughput is defined as the product of the unconditioned success probability and the sum rate per unit area assuming that capacity-achieving codes are used. 
When $K$ independent data streams are sent on each broadcast cluster the throughput is given by
\begin{equation}
\mathcal{T}  =  \lambda\sum_{k \in \mathcal{K}}\mathbb{P}\left(\sinr_k > \beta_k \right)\log_2(1+\beta_k).
\label{thruput}
\end{equation}
Note that the success probability $\mathbb{P}\left(\sinr_k > \beta_k \right)$ is itself a monotonically decreasing function of $\lambda$. As the success probability is not constrained to a minimum value, the throughput-maximizing density may be obtained at the expense of very high outage levels.

\emph{Multi-stream transmission capacity.} Generalizing \cite{WebYan05} for the case where $K$ streams are sent by each source node, we define the multi-stream transmission capacity (TC) as the maximum number of concurrent multi-stream transmissions $\lambda_{\epsilon}$ per unit area allowed subject to an outage constraint $\epsilon$, i.e.
\begin{equation}
\mathcal{C}  =  K\lambda_\epsilon(1-\epsilon),\label{multicapa} 
\end{equation}
\begin{equation}
\hspace{-40mm} \text{where} \hspace{45mm} \lambda_\epsilon = \sup \left\{\lambda : \mathbb{P}\left(\sinr_k \leq \beta_k \right) \leq \epsilon_k, \ \forall k \in \mathcal{K} \right\}
\end{equation}
defines the maximum contention density for a per-stream outage constraint $\epsilon_k \in (0,1)$. This outage-based metric quantifies how efficiently the network utilizes space as resource under a maximum outage constraint, as opposed to the network throughput that may result in high outage events.
In other words, it calculates the maximum density of transmissions per unit area so that all $K$ users in the broadcast cluster do not exceed a desired outage level $\epsilon$.


\emph{Average ergodic rate.} Finally, we define the average data rate (in nats/Hz) achievable by a typical user assuming Shannon capacity achieving modulation and coding for the instantaneous SINR to be
\begin{equation}
\label{ergC}
\mathcal{R}(\lambda,\alpha) = \mathbb{E}\left\{\log(1+\sinr)\right\}.
\end{equation}
In contrast to the two aforementioned metrics, this average capacity measure presumes dynamic rate adaptation to the instantaneous SINR.

In the remainder, for the sake of exposition simplicity we assume that all streams have identical stream outage constraint $\epsilon$ and SINR target $\beta$, i.e. $\epsilon_k = \epsilon$ and $\beta_k=\beta$, $\forall k\in\mathcal{K}$.

\section{ZFBF Performance Analysis}
\label{sec: ZFperf}
In this section, we derive new closed-forms expressions for the network throughput, transmission capacity, and average achievable rate of zero-forcing precoding.

\subsection{Outage Probability}

\begin{theorem}
\label{Th:Pout}
\textit{The outage probability for the $k$-th typical user in a wireless ad hoc network using multiuser zero-forcing with quantized CSIT is given by
\begin{equation*}
\pkout = 1 - \frac{e^{-\lambda\mathcal{I}_{K}\zeta_k^{2/\alpha}}e^{-\sigma^2\zeta_k/\rho}}{(1+\beta_k\delta)^{K-1}}
\label{eq:pout}
\end{equation*}
where $\zeta_k = \beta_k R_k^{\alpha}$, $\delta = 2^{-\frac{B}{M-1}}$, and $\mathcal{I}_K = \frac{2\pi}{\alpha}\displaystyle \sum_{m=0}^{K-1}\binom{K}{m}B\left(m+\frac{2}{\alpha},K-m-\frac{2}{\alpha}\right)$,\\
with $B(a,b) = \int_{0}^{1}t^{a-1}(1-t)^{b-1}dt = \frac{\Gamma(a)\Gamma(b)}{\Gamma(a+b)}$ being the Beta function and $\Gamma(x) = \int_{0}^{\infty}t^{x-1}e^{-t}dt$ the Gamma function.}
\end{theorem}
\IEEEproof{See Appendix~\ref{Th:Pout_proof}.}

\begin{figure}[ht]
\centering
\includegraphics[width=4.7in]{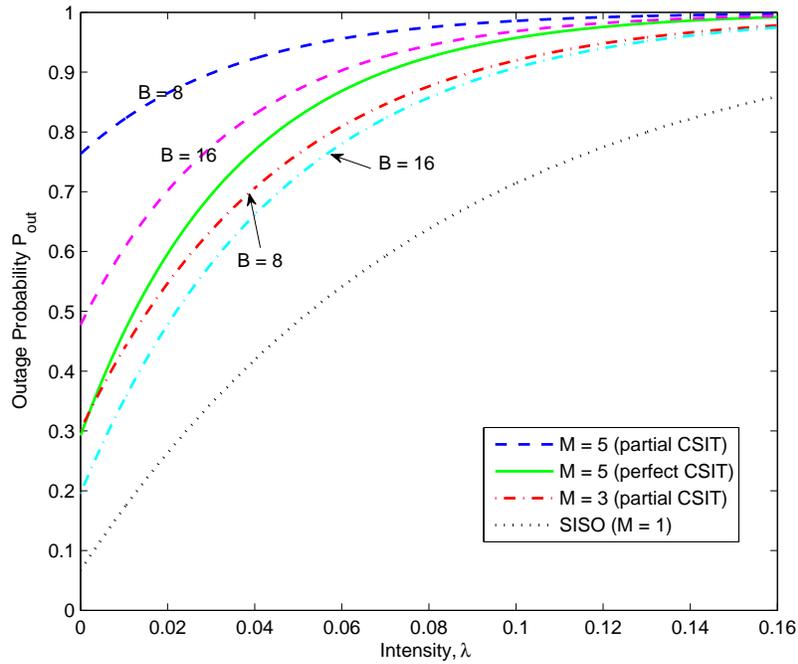}
\vspace{-5mm}
\caption{Outage probability vs. node density at SNR = 20 dB, for $\alpha = 4$, $d = 1.5$, and $\beta = 1$ dB.}
\label{fig1}
\end{figure}

As expected, the outage probability is a decreasing function with the feedback bit rate $B$ and an increasing function of $K$ since $\frac{\partial {\mathcal{F}^{(k)}}}{\partial K} > 0$. Fig.~\ref{fig1} shows the outage probability vs. the node intensity for different values of feedback load and antennas. The single-antenna, single-stream (SISO) outage probability is also plotted for comparison. Numerical evaluations of Theorem~\ref{Th:Pout} confirm the analysis and show that increasing the number of antennas/streams results in unacceptably high outage probability values even in sparse networks ($\lambda \to 0$). 

\subsection{Network Throughput}
Based on the above derived outage probability and the throughput definition (cf.(\ref{thruput})), when $K$ independent data streams are sent on each broadcast cluster, the total number of successful bits/s/Hz/unit area (throughput) is given by
\begin{eqnarray}
\mathcal{T} = \lambda\sum_{k \in \mathcal{K}}\frac{e^{-\lambda\mathcal{I}_{K}\zeta_k^{2/\alpha}}e^{-\sigma^2\zeta_k/\rho}}{(1+\beta_k\delta)^{K-1}}\log_2(1+\beta_k) \stackrel{(a)} \geq  K\lambda \frac{e^{-\lambda\mathcal{I}_{K}\zeta_{\rm max}^{2/\alpha}}e^{-\frac{\sigma^2\zeta_{\rm max}}{\rho}}}{(1+\beta\delta)^{K-1}}\log_2(1+\beta).
\label{thruput2}
\end{eqnarray}
where (a) results by setting $\beta_k=\beta$, $\forall k\in\mathcal{K}$, and $\zeta_{\rm max} = \beta d_{\rm max}^{\alpha}$ with $d_{\rm max} = \displaystyle \max_{k \in \mathcal{K}} d_{k}$. The approximation is derived for the sake of exposition simplicity and can be seen as a lower bound on the network throughput with equal target SINR for all users. 
In the sequel, for exposition convenience and unless otherwise stated, we consider that $K=M$ users are served on each cluster.  

\begin{remark}
By taking the derivative of $\mathcal{T}$ with respect to $\lambda$ keeping all other parameters fixed, we see that throughput decreases with the node density, if
\begin{equation}
\lambda \geq (\mathcal{I}_{M} \beta^{\frac{2}{\alpha}}d_{\rm max}^2)^{-1} = \lambda^*. 
\end{equation}
The optimal intensity $\lambda^*$ does not depend on the number of feedback bits, which only affects the amount of inter-cluster interference. However, although throughput can be maximized for $\lambda \geq \lambda^*$, the number of outage events can be arbitrarily high due to intra-cluster interference. For $\lambda = \lambda^*$, the success probability becomes $\frac{1}{e}\frac{e^{-\sigma^2\zeta_{\rm max}/\rho}}{(1+\beta\delta)^{M-1}}$, which means that multiuser zero-forcing beamforming with finite rate feedback decreases the success probability by a factor of $(1+\beta\delta)^{M-1}$ as compared to the case of point-to-point ad hoc communications.
Furthermore, $\lambda^*$ is a regularly varying function of $M$ with index $-2/\alpha$ and for large $M$, the optimal network density scales as $\lambda^* = O(M^{-\frac{2}{\alpha}})$, namely $\displaystyle \lim_{M \to \infty} \frac{\lambda^*}{M^{-\frac{2}{\alpha}}} = \left(\pi \beta^{\frac{2}{\alpha}}d_{\rm max}^2 \Gamma(1-2/\alpha)\right)^{-1}$.
As expected, the optimal contention density decreases when $M$ increases since increasing the number of streams sent boost the inter-cluster interference $I_p$. The optimal density also decreases for $\beta$ or $d_{\rm max}$ increasing as the reliability requirements on the per-user performance become higher and harder to satisfy.
\end{remark}

\begin{remark}
In terms of feedback rate, the network throughput can be shown to be a monotonically increasing function with $B$. Focusing now on the transmit antenna configuration, after some algebraic manipulations, we can show that throughput is maximized if 
\begin{equation*}
M^* = \max\left(\left\lfloor \ell\right\rfloor,1\right) 
\end{equation*}
where $\ell$ is the nontrivial solution for $M$ of 
\begin{equation*}
\frac{M}{M-1}\cdot(\log2)\cdot\frac{\beta\delta}{1+\beta\delta}B + \frac{\sigma^2\zeta_{\rm max}}{\rho} + \lambda M\zeta_{\rm max}^{2/\alpha}\frac{\partial \mathcal{I}_{M}}{\partial M} = 1.
\end{equation*}
Although the partial derivative of $\mathcal{I}_{M}$ can be expressed as sum of beta and digamma functions $\psi_0(x)$ since $\frac{\partial \rm B(y,x)}{\partial x} = B(y,x)(\psi_0(x)-\psi_0(x+y))$, a closed-form expression for $\ell$ is hard to obtain.\\ 
An analytical expression for $M^*$ can be found by applying the large $M$ approximation $\mathcal{I}_M \sim \pi\Gamma(1-2/\alpha)M^{2/\alpha}$.
In that case $M^* = x^{\alpha}$, where $x$ is the solution of the polynomial equation $c_3x^{2\alpha} + c_2x^{\alpha+2} + (c_1-c_3-1)x^{\alpha}- c_2x^2 + 1 = 0$, where $c_1 = \frac{B\beta\delta}{1+\beta\delta}\log2$, $c_2 = \lambda\pi\Gamma(1-2/\alpha)\zeta_{\rm max}^{2/\alpha}$, and $c_3 = \frac{\sigma^2\zeta_{\rm max}}{P}$. From Abel's impossibility theorem~\cite{Algebrabook}, a formula solution only exists for $a \leq 4$, while for $a = 3$ the solution can be expressed using Kamp\'e de F\'eriet functions. 
\end{remark}

\begin{figure}[ht]
\centering
\includegraphics[width=5in]{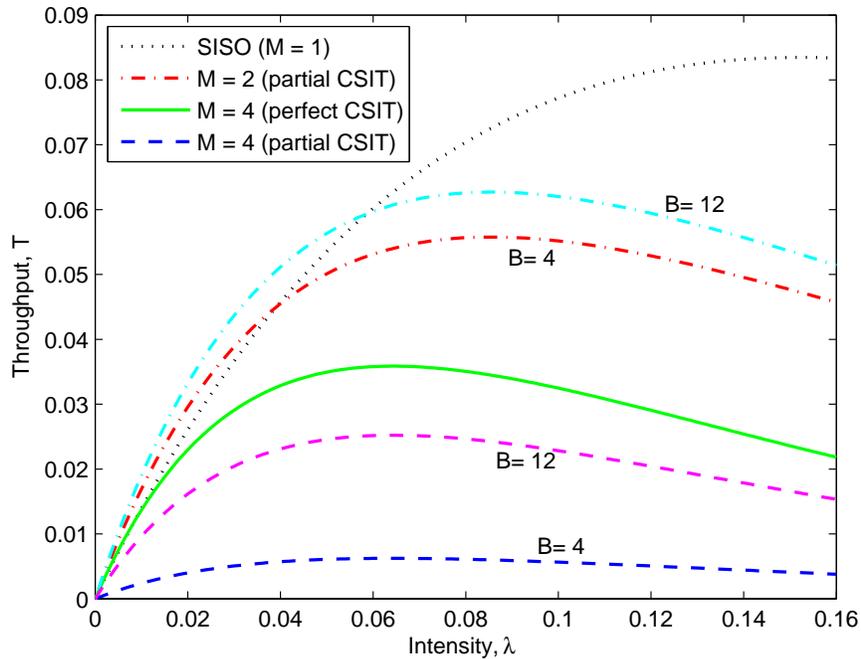}
\vspace{-5mm}
\caption{Throughput vs. node density for $\alpha = 4.2$, $d = 1.5$, $\beta = 3$ dB, and SNR = 15 dB.}
\label{fig3}
\end{figure}

In Fig. \ref{fig3} we plot the network throughput vs. the intensity $\lambda$. We observe that throughput is \emph{a decreasing function of the number of antennas} and that the performance degradation from imperfect CSIT is more pronounced for $M$ increasing. The SISO and the perfect CSI-based zero-forcing are also plotted for comparison. We also see that multi-stream transmission is slightly superior in sparse networks (low $\lambda$), but is generally outperformed by SISO.

\subsection{Multi-stream Transmission Capacity}
We turn now our attention to the maximum achievable throughput under bounded outage levels. 

\begin{theorem}
\textit{The maximum multi-stream transmission capacity of limited feedback zero-forcing precoding in random access ad hoc networks is given by
\begin{equation}
\mathcal{C} = \frac{K(1-\epsilon)}{\mathcal{I}_{K}\zeta_{\rm max}^{2/\alpha}}\left(\log\frac{1}{1-\epsilon} - \frac{\sigma^2\zeta_{\rm max}}{\rho}- \log(1+\beta\delta)^{K-1}\right).
\label{eq:TC}
\end{equation}}
\end{theorem}
\begin{proof}
The result follows by finding the inverse of the expression $\mathcal{F}_m(\beta, \alpha) = \mathbb{P}\left(\displaystyle \min_{k \in \mathcal{K}}\sinr_k \leq \beta \right)$ $= \epsilon$ with respect to $\lambda$, i.e. $\mathcal{F}_m^{-1}(\beta, \alpha)$, and substituting it in (\ref{multicapa}).
\end{proof}

The second term in (\ref{eq:TC}) captures the effect of background noise on multi-stream TC, whereas the third term corresponds to the capacity degradation from quantized CSIT. 
For large $M$ with $K=M$, the first term scales as $\Theta(M^{1-\frac{2}{\alpha}})$, whereas both second and third terms scale as $\Theta(M^{2-\frac{2}{\alpha}})$ (for fixed feedback quality). This implies that the detrimental effect of residual interference from quantized channel information is orderwise dominant, becoming the transmission capacity limiting factor.

Interestingly, in contrast to the cases of point-to-point and multiuser ad hoc communication with perfect CSI, it is not guaranteed that non-zero transmission capacity can be achieved for any feedback rate due to the self-interference that cannot be completely eliminated with quantized CSIT. After some algebra, we can show that the amount of feedback resolution $B_f$ defining the multi-stream TC feasibility region, i.e. the region for which positive maximum contention density $\lambda_{\epsilon}$ exists, is
\begin{equation}
B_f > \left\lceil (M-1)\log_2\left(\frac{\beta}{(\frac{e^{-\sigma^2\zeta_{\rm max}/\rho}}{1-\epsilon})^{\frac{1}{K-1}}-1}\right)\right\rceil
\label{FB_TCpositive}
\end{equation}
provided that $(1-\epsilon)e^{\sigma^2\zeta_{\rm max}/\rho} < 1$. The latter condition is more general and applies even to systems with perfect CSI as it guarantees that a non negative $\lambda$ exists for certain $\snr = \rho/\sigma^2$ and outage constraints.

\begin{figure}[ht]
\centering
\includegraphics[width=5in]{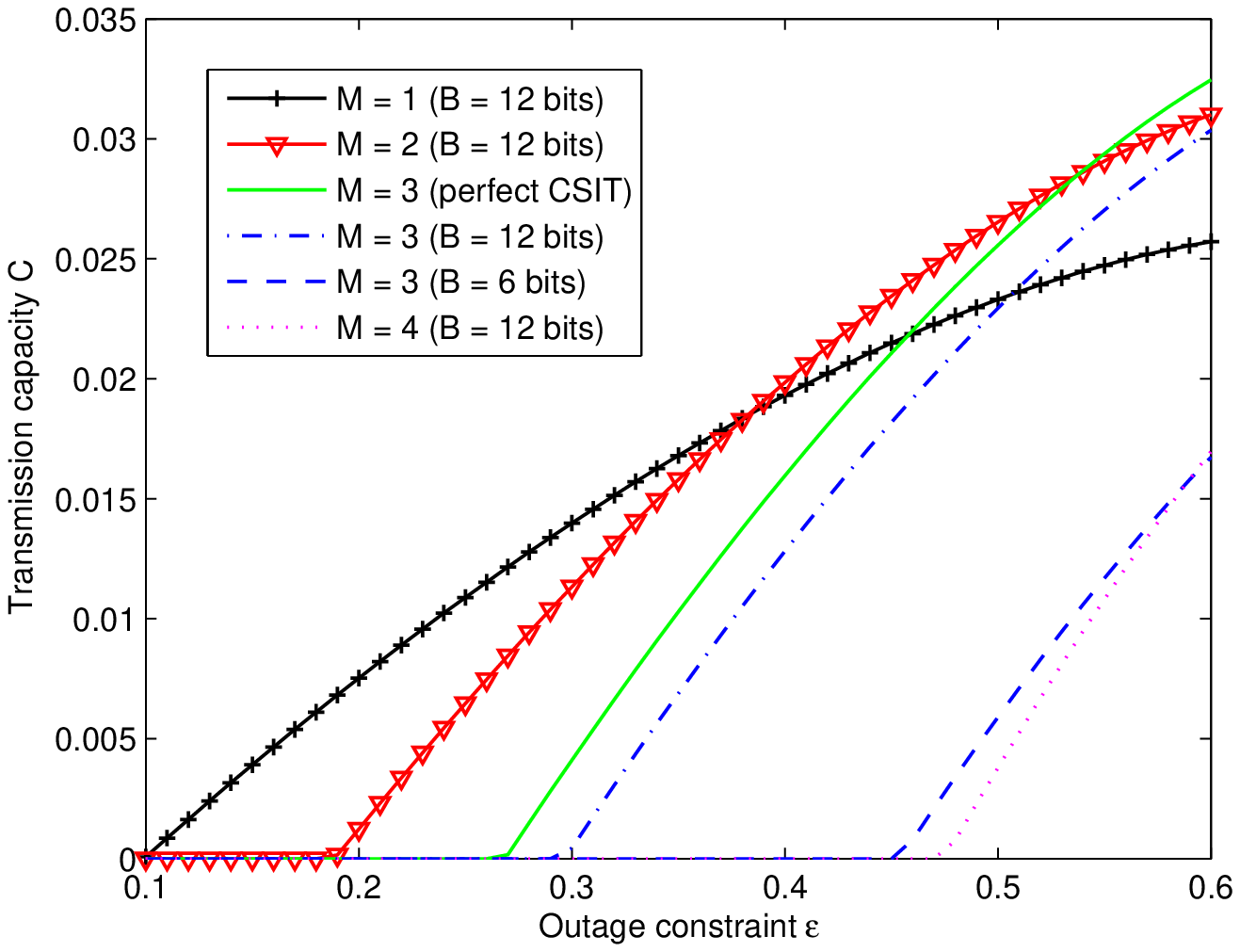}
\caption{Transmission capacity vs. outage constraint for $\alpha = 4.5$, $\beta = 1$ dB, and SNR = 20 dB.}
\label{fig5}
\end{figure}

\begin{figure}[ht]
\centering
\includegraphics[width=4.8in]{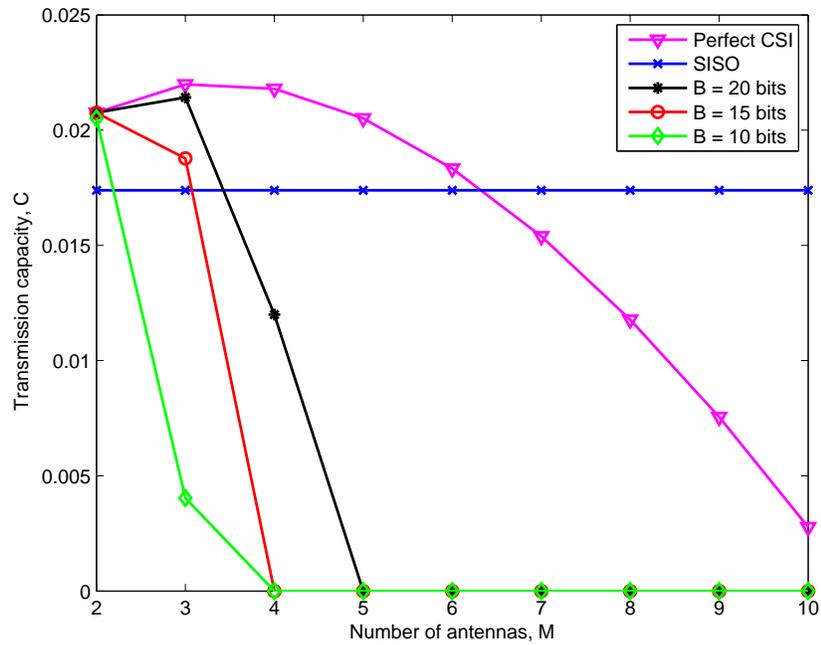}
\vspace{-5mm}
\caption{Transmission capacity vs. number of antennas/streams for $\alpha = 4$, $\epsilon = 0.1$, $d = 1$, $\beta = 0$ dB, and SNR = 20 dB.}
\label{fig6}
\end{figure}

In Figs. \ref{fig5} and \ref{fig6} we plot the transmission capacity vs. the outage constraint and the number of antennas, respectively. We observe that positive transmission capacity is achieved for significantly high outage $\epsilon$, while increasing the number of streams further deteriorates the performance. In practically relevant scenarios where the outage constraint is kept low, single-stream transmission (TDMA) is optimal, even in the high resolution regime ($B \to \infty$). Furthermore, we see that multi-stream transmission is beneficial for low number of antennas/streams and for relatively large number of feedback bits.

\subsection{Average Achievable Rate}
In this section, we derive the average data rate achievable by a typical receiver assuming Shannon capacity achieving modulation and coding for the instantaneous SINR. Note that this is not the maximum achievable Shannon capacity as each transmitter-receiver link treats interference as noise. 
\begin{theorem}
\label{Th_Erg}
\textit{The ergodic rate in nat/s/Hz of a typical receiver in a broadcast cluster where multipacket transmission is employed using finite-rate feedback zero-forcing is given by
\begin{eqnarray}
\label{rate_intg}
\mathcal{R}_k  & = &  \mathbb{E}\left\{\log(1+\sinr)\right\}
 =  \int_{0}^{\infty}\frac{e^{-C_1x}e^{-C_2x^{\frac{2}{\alpha}}}}{(1+x)(1+\delta x)^{M-1}}\rm{d}x.
\end{eqnarray}
where $C_1 = \sigma^2R_k^{\alpha}/\rho$ and $C_2 = \lambda\mathcal{I}_{M}R_k^{2}$. 
}
\end{theorem}
\IEEEproof{See Appendix~\ref{Th_Erg_proof}.}

For general values of $\alpha > 2$, the computation of the average user rate $\mathcal{R}_k$ involves numerical integration. In the interference-limited regime ($\sigma^2 \to 0$), pseudo-closed-form expressions involving generalized hypergeometric functions can be found, but these provide little insight on how different system operating parameters affect the average user rate. Therefore, we consider concise performance bounds.

As the main focus of the paper is to investigate the relationship among feedback bit rate $B$ and both inter-cluster and intra-cluster interference, we first provide the following result that shows that the average achievable rate with limited feedback of fixed quality converges to a finite ceiling as $\rm SNR \to \infty$.
\begin{theorem}
\label{Th_ErgJ}
\textit{The average user rate of imperfect CSIT-based zero-forcing is upper bounded by
\begin{eqnarray}
\label{eq:th4}
\mathcal{R}_k  \leq \frac{B\log2}{M-1} + H_{M-1} - \psi(M) + \log\left(1 + d_k^{2\alpha}\left(\frac{\pi\lambda}{\left\lceil \frac{\alpha}{2}\right\rceil-1}\right)^{\frac{\alpha}{2}} + \delta(M-1) + \frac{d_k^{\alpha}\sigma^2}{\rho}\right)
\end{eqnarray}
where $H_{n} = \sum_{i=1}^{n}\frac{1}{i}$ is the $n$-th harmonic number and $\psi(\cdot)$ is the digamma function. 
}
\end{theorem}
\IEEEproof{See Appendix~\ref{Th_ErgJ_proof}.}

The above result implies that at high SNR ($P \to \infty$), the user rate is bounded and the system becomes interference-limited no matter how many feedback bits are reported back to the transmitter. The upper bound in the above theorem is quite loose in general, however it was derived for demonstrating the quasi linear dependence of the average rate and the feedback load $B$.
We derive now a tighter upper bound by applying integral inequalities directly to (\ref{rate_intg}) as a means to find closed-form expression for the mean ergodic rate.
\begin{lemma}
\label{Holder_Erg}
\textit{The average achievable user rate is upper bounded by
\begin{eqnarray}
\label{rate_integral}
\mathcal{R}_k & \leq & \displaystyle \min_{u_1,v_1} \left(\int_{0}^{\infty}e^{-u_1C_1x}e^{-u_1C_2x^{\frac{2}{\alpha}}}\rm{d}x \right)^{\frac{1}{u_1}} \cdot (\mathcal{A}(v_1))^{\frac{1}{v_1}} \label{hold1}\\
& \leq & \displaystyle \min_{u_1,v_1,u_2,v_2} (u_1u_2C_1)^{-\frac{1}{u_1u_2}}(\Gamma(1+\alpha/2))^{\frac{1}{u_1v_2}}(u_1v_2C_2)^{-\frac{\alpha}{2u_1v_2}}\cdot \left[\mathcal{A}(v_1)\right]^{\frac{1}{v_1}}, \label{hold2} 
\end{eqnarray}
with $1< u_1,u_2,v_1,v_2<\infty$, $\frac{1}{u_1} + \frac{1}{v_1} = 1$, $\frac{1}{u_2} + \frac{1}{v_2} = 1$, and
\begin{eqnarray}
\label{holder_I}
\mathcal{A}(v_1) = (\delta-1)^{1-Mv_1}\delta^{v_1-1}B(Mv_1-1,1-(M-1)v_1) + \frac{_2F_1(1,v_1,2-(M-1)v_1,1/\delta)}{v_1(M-1)\delta}
\end{eqnarray}
where $_2F_1(a,b,c,z)$ denotes the $_2F_1$ Gauss hypergeometric function, and $B(x,y)$ the Beta function. 
}
\end{lemma}
\IEEEproof{See Appendix~\ref{Holder_Erg_proof}.}

\begin{figure}[ht]
\centering
\includegraphics[width=4.8in]{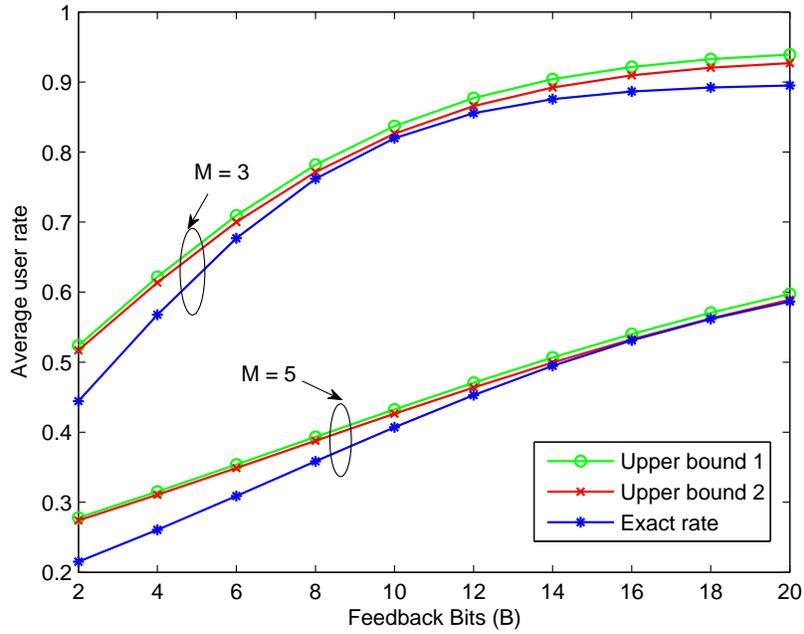}
\vspace{-5mm}
\caption{Average rate vs. feedback bits for $\alpha = 3.8$, $d = 1$, $\lambda = 0.05$, and SNR = 20 dB.}
\label{fig7}
\end{figure}

For $u_1=u_2=v_1=v_2=2$, which provides the tightest upper bound in most cases, significant simplification is possible for (\ref{holder_I}) since $\mathcal{A}(2)$ is given in terms of $\log(\delta)$ and a polynomial expression in $\delta$. In that case, we can easily show that the upper bound on ergodic capacity scales like $\Theta(\lambda^{-\alpha/8})$. The validity of the above bounds is verified in Fig. \ref{fig7}, where the ergodic capacity given by (\ref{hold1}) (cf. upper bound 1) and (\ref{hold2}) (cf. upper bound 2) is compared with the exact average user rate for different number of feedback bits. We observe that the tightness of our bounds is improved when the number of antennas and feedback bits is increased.

In the no noise case, the above lemma results in
\begin{eqnarray}
\mathcal{R}_k \leq \frac{\left[\Gamma\left(1+\frac{\alpha}{2}\right)\right]^{\frac{1}{\alpha}}}{d_k\sqrt{\alpha \lambda \mathcal{I}_{\rm M}}} \left(\int_{0}^{\infty}\left[\frac{(1+\delta x)^{1-M}}{1+x}\right]^{\frac{\alpha}{\alpha-1}}\textrm{d}x\right)^{1-\frac{1}{\alpha}},
\label{eq:holder_icc}
\end{eqnarray}
which scales as $\Theta(\lambda^{-\frac{1}{2}})$ and is inversely proportional to the distance between transmitter and the $k$-th user. For $\lambda$ and $B$ that do not depend on $M$ and for large $M$, by calculating the integral, the upper bound in (\ref{eq:holder_icc}) is shown to be inversely proportional to $M$. This implies that in order to have per-user mean rate that does not scale with the number of antennas, $\lambda \sim M^{-2}$.
Finally, in the interference-limited regime and for $\alpha = 4$, (\ref{rate_intg}) admits a closed-form expression in terms of Meijer-G functions and trigonometric integrals. 
The no noise upper bound (\ref{eq:holder_icc}) approximates very well the exact ergodic rate even for moderate SNR values for increasing number of antennas. 

Returning now to the general case, the following easily computable result can be obtained using different bounding techniques.
\begin{lemma}
\label{Bernoulli_Erg}
\textit{The average user rate with finite rate-based zero-forcing satisfies
\begin{eqnarray}
\label{rate_bernoul_bound}
\mathcal{R}_k^{\rm LB} \leq \mathcal{R}_k \leq \mathcal{R}_k^{\rm UB}
\end{eqnarray}
with 
\begin{eqnarray}
\mathcal{R}_k^{\rm LB} & = & -2\cos(C_2){\rm Ci}(C_2) + \frac{\delta(1-M)}{\sqrt{\pi}}\MeijerG{-1}{}{-1,0,\frac{1}{2}}{}{\frac{C_2^2}{4}} + \sin(C_2)(\pi - 2{\rm Si}(C_2)) \label{bern1}\\ 
\mathcal{R}_k^{\rm UB} & = & \frac{1}{(M-1)\delta-1}[2\cos(C_2){\rm Ci}(C_2) - 2\cos(\tilde{C_2}) {\rm Ci}(\tilde{C_2}) \\
& - &\pi \sin(C_2) + \pi \sin(\tilde{C_2}) + 2\sin(C_2){\rm Si}(C_2) - 2\sin(\tilde{C_2}) {\rm Si}(\tilde{C_2})] \label{bern2}
\end{eqnarray}
where $\MeijerG[a,b]{n}{p}{m}{q}{z}$ is the Meijer-G function, ${\rm Si}(x) = \int_{0}^{x}\frac{\sin t}{t}dt$ is the sine integral, ${\rm Ci}(x) = - \int_{x}^{\infty}\frac{\cos t}{t}dt$ is the cosine integral, and $\tilde{C_2}= \frac{C_2}{(M-1)\delta}$.}
\end{lemma}
\begin{proof}
The bounds are obtained by applying Bernoulli's inequality \cite{HardyIneq} for the integrand and evaluating the resulting integrals. Specifically, for the lower bound, we apply the inequality $(1+x)^{r} \geq 1+rx$, for $x > -1$, $r \leq 0 \ \text{or} \ r \geq 1$, to the function $g(x) = (1+\delta x)^{M-1}$. For the upper bound, we use that $(1+x)^{-r} \leq (1+rx)^{-1}$ for $r > 0, x \geq - 1$.
\end{proof}

\begin{figure}[ht]
\centering
\includegraphics[width=4.8in]{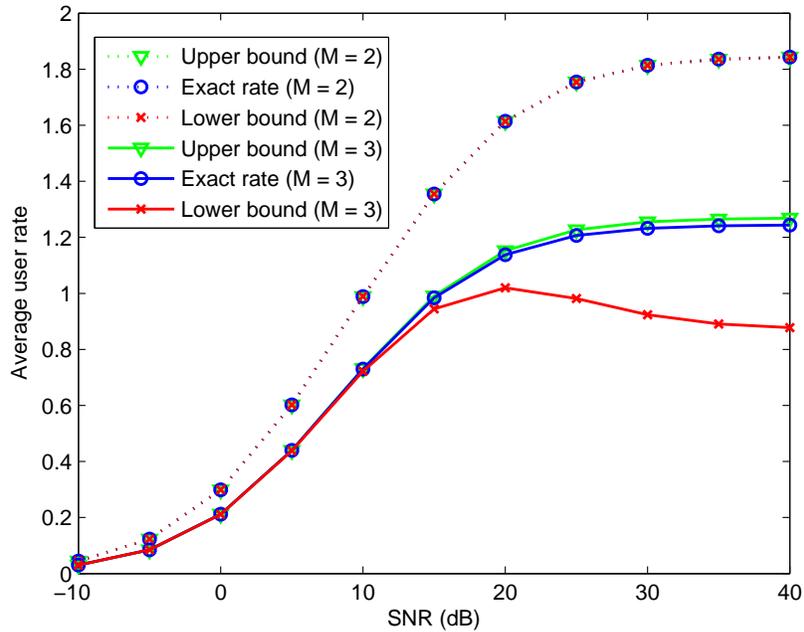}
\vspace{-5mm}
\caption{Average rate vs. SNR for $\alpha = 4.2$, $d = 1$, $\lambda = 0.05$, $M = 3$, and $B = 10$ bits.}
\label{fig9}
\end{figure}

In Fig. \ref{fig9}, the achievable mean user rate is compared with (\ref{bern2}) (cf. upper bound) and (\ref{bern1}) (cf. lower bound) vs. SNR. We observe that the bounds are very tight at low $M$ for all SNR range, while the lower bound becomes loose at high SNR when the number of antennas is increased.


\section{Effect of Limited Feedback}
\label{sec:LFBeffect}
In this section, we analyze the effect of feedback quality on the network performance and provide design guidelines for the system operating points based on our analytical framework. In particular, we show at which rate feedback has to scale to maintain a certain bounded network capacity gap. We also derive the optimal number of streams/users to be employed in order to maximize the network throughput and the multi-stream transmission capacity.

\subsection{Performance Degradation due to Finite Rate Feedback}
We first provide the feedback bit scaling that guarantees constant (bounded) performance loss between the performance of zero-forcing with perfect CSI and that with partial CSIT. 

\subsubsection{Transmission Capacity} 
The transmission capacity gap $\Delta\mathcal{C}$ is defined as the difference between the transmission capacity achieved by perfect CSIT-based and that of limited feedback-based zero-forcing, i.e. $\Delta\mathcal{C} = \left(\mathcal{C}_{\rm CSI} - \mathcal{C}\right)$, where $\mathcal{C}_{\rm CSI}$ is the multi-stream transmission capacity given by (\ref{eq:TC}) for $B \to \infty$ (perfect CSI). Thus, the performance degradation is given by
\begin{equation*}
\Delta\mathcal{C} = \frac{K(1-\epsilon)}{\mathcal{I}_{K}\zeta_{\rm max}^{2/\alpha}}\log(1+\beta\delta)^{K-1}.
\end{equation*}
In order to maintain a transmission capacity offset $\Delta\mathcal{C} = \log c$, after some algebraic manipulations, we have that the number of feedback bits per user satisfies
\begin{equation*}
B_{\Delta\mathcal{C}} \geq (M-1)\log_2\beta - (M-1)\log_2\left(c^{\frac{\mathcal{I}_{K}\zeta_{\rm max}^{2/\alpha}}{K(K-1)(1-\epsilon)}} - 1\right) \ \ \ \text{bits/user,}
\label{DTC:bits}
\end{equation*}
with $1 \leq c \leq (1+\beta)^{\frac{K(K-1)(1-\epsilon)}{\mathcal{I}_{K}\zeta_{\rm max}^{2/\alpha}}}$ for a non trivial result. 
Therefore, to guarantee a constant performance offset in terms of transmission capacity, the number of feedback bits per user must be increased at least linearly with the number of antennas/streams and approximately logarithmically with the target SINR constraint. This is basically the same scaling behavior as in \cite{Jindal_finiteFB} for fixed average rate offset if the target SINR constraint is interchanged with the transmit power. Interestingly, TC for ad hoc networks appears to capture the performance degradation due to multiuser interference similar to ergodic capacity for single-cell systems.

\subsubsection{Network Throughput}
Define now the network throughput ratio gap $\mathcal{Q}\mathcal{T}$ to be the ratio of the throughput achieved by zero-forcing with perfect CSI to the throughput of finite rate feedback zero-forcing, which is given by $\mathcal{Q}\mathcal{T} = (1+\beta\delta)^{K-1}$.
The number of feedback bits per user for throughput ratio offset  $\mathcal{Q}\mathcal{T} = r$ needs to scale according to
\begin{equation*}
B_{\mathcal{Q}\mathcal{T}} \geq (M-1)\log_2\beta - (M-1)\log_2\left(r^{\frac{1}{K-1}} - 1\right)\ \ \ \text{bits/user,}
\end{equation*}
for any $r$ satisfying $1 \leq r \leq (1+\beta)^{K-1}$. 

Similarly to the transmission capacity offset, the number of feedback bits has to increase at least linearly with the number of antennas/streams and logarithmically with the target SINR constraint to maintain constant throughput loss. For instance, for $\beta = 3$ dB and $M = K = 4$, at least $B = 9$ bits are required for a 3-dB offset. Furthermore, for $r = 2^{K-1}$, the resulting feedback scaling takes on the following simple form
\begin{equation} 
B_{\mathcal{Q}\mathcal{T}} = \frac{M-1}{3}\beta_{\rm dB} \ \ \ \text{bits/user.}
\end{equation} 
Note that the feedback bit scaling takes on identical form with that in \cite{Jindal_finiteFB} for a 3-dB rate offset substituting transmit power $P$ for target SINR $\beta$.

\subsection{Feedback Rate Scaling}
In the previous section, we quantified the number of feedback bits required for a fixed gap degradation in the throughput and transmission capacity performance of zero-forcing with imperfect feedback. Here, we provide design guidelines on the feedback bit scaling for asymptotically vanishing performance loss due to partial CSIT.

First, similar to MISO broadcast channels without inter-cell interference, both throughput and transmission capacity of zero-forcing with limited feedback are bounded with fixed $B$ even if other system parameters grow large. Furthermore, if the feedback bits do not scale with $M$ and/or $\beta$, the throughput ratio and the transmission capacity offset become unbounded for asymptotically large values of $M$, $\beta$. 

In the high antenna/stream regime, it can be shown that if the feedback load $B$ is scaled with $M$ at a rate strictly greater than $(M-1)\log_2M$, i.e., $B = (M-1)\log_2(M^{\eta})$ for any $\eta > 1$, the transmission capacity offset converges to zero, i.e. $\displaystyle \lim_{M \to \infty}\Delta\mathcal{C} = \displaystyle \lim_{M \to \infty}\frac{(1-\epsilon)}{\mathcal{I}_{M}\zeta_{\rm max}^{2/\alpha}}\log(1+\beta\delta)^{M-1} = 0$, and the throughput ratio gap converges to one, i.e. $\displaystyle \lim_{M \to \infty}\mathcal{Q}\mathcal{T} = \displaystyle \lim_{M \to \infty}(1+\beta\delta)^{M-1} \to 1$.

The rate of convergence to zero and one respectively depends on $\eta$ and is faster with $\eta$ increasing. The throughput ratio converges to one also in the case where $B$ scales superlinearly with the number of antennas, i.e. $B = M^{\eta}$. Based on the above results, we establish that, at asymptotically high $M$ and under the aforementioned bit scaling, the network throughput and the transmission capacity of finite rate feedback zero-forcing converges to the perfect CSI throughput. In contrast, if the feedback rate is not properly adapted, the transmission capacity offset scales as $\Delta\mathcal{C} = O(M^{1-2/\alpha})$. 

In the high reliability regime (large $\beta$), if the feedback load $B$ is scaled with $\beta$ at a rate strictly greater than $(M-1)\log_2\beta$, i.e. $B = \kappa\log_2\beta$ for any $\kappa > M-1$, the throughput ratio gap converges to one, i.e. $\displaystyle \lim_{\beta \to \infty}\mathcal{Q}\mathcal{T} = \displaystyle \lim_{\beta \to \infty}(1+\beta\delta)^{M-1} \to 1$.

Under the same bit scaling, the transmission capacity offset vanishes (asymptotically in $\beta$), i.e. $\displaystyle \lim_{M \to \infty}\Delta\mathcal{C} = 0$.
Thus, at large $\beta$ (high SINR regime) and under the aforementioned bit scaling, the throughput of the finite rate feedback ZFBF converges weakly to the perfect CSI throughput.

\subsection{Optimal Number of Streams}
In this section, we investigate the optimal number of streams to be used per cluster in order to maximize the capacity. These results also provide useful insights on the feasibility and the potential gains of multi-stream, multiuser beamforming and adaptive beam selection in wireless ad hoc networks with imperfect feedback. 

For that, we consider again that $K \leq M$ streams can be sent. 
We define the per-user throughput as the normalized network throughput over the number of users, i.e. $\mathcal{T}_{u} = \frac{1}{K}\mathcal{T}$. Taking the partial derivative with respect to the number of streams, we can show that $\mathcal{T}_{u}$ is a decreasing function with $K$.
This confirms the intuitive argument that, from the user perspective, employing single-stream beamforming ($K^* = 1$) maximizes the per-user throughput.  

However, the optimal number of users to serve may alter if we consider the system (broadcast cluster) overall throughput. In this case, there is a tradeoff between spatial reuse and feedback quality, i.e. for certain values of $K$ and other system parameters, the spatial multiplexing gain may compensate for the performance degradation incurred due to finite rate feedback.  
For the network throughput, the complicated form of the interference constant $\mathcal{I}_K$ precludes a simple, closed-form expression for the optimal number of streams. Specifically, we have the following result:
\begin{prop}
\textit{The number of users to be served per broadcast cluster that maximizes the network throughput is
\begin{equation*}
K^* = \min\left(\max\left(\left\lceil \omega \right\rceil,1\right),M\right) 
\end{equation*}
where $\omega$ is the solution for $K$ of 
\begin{equation*}
K\left(\lambda\zeta_{\rm max}^{2/\alpha}\frac{\partial \mathcal{I}_K}{\partial K} + \frac{\sigma^2\zeta_{\rm max}}{P} + \log(1+\beta\delta) \right)  = 1.
\end{equation*}}
\end{prop}
\begin{proof}
The result following by taking the derivative of (\ref{thruput2}) with respect to $K$ and finding the optimal value of $K$ based on Fermat's theorem for the stationary points.
\end{proof}

\begin{remark}
Using the large $K$ approximation $\mathcal{I}_K \sim \pi\Gamma(1-2/\alpha)K^{\frac{2}{\alpha}}$, we have that $K^* = x^{\alpha}$, where $x$ is the solution of the polynomial equation $c_1x^{\alpha} + c_2x^2 - 1 = 0$, where $c_1 = \frac{\sigma^2\zeta_{\rm max}}{P} + \log(1+\beta\delta)$ and $c_2 = \lambda\pi\Gamma(1-2/\alpha)\zeta_{\rm max}^{2/\alpha}$. By the Abel-Ruffini theorem, no general algebraic solution exists for $a \geq 5$, however for this particular form of the polynomial equation, $K^*$ can be found in closed-form for $\alpha \in \{2,3,4,5,6,8\}$; solutions for $\alpha = 6$ and $\alpha = 8$ will be the squared root of the $\alpha = 3$ and $\alpha = 4$ solutions, respectively. 
\end{remark}

\begin{figure}[ht]
\centering
\includegraphics[width=4.9in]{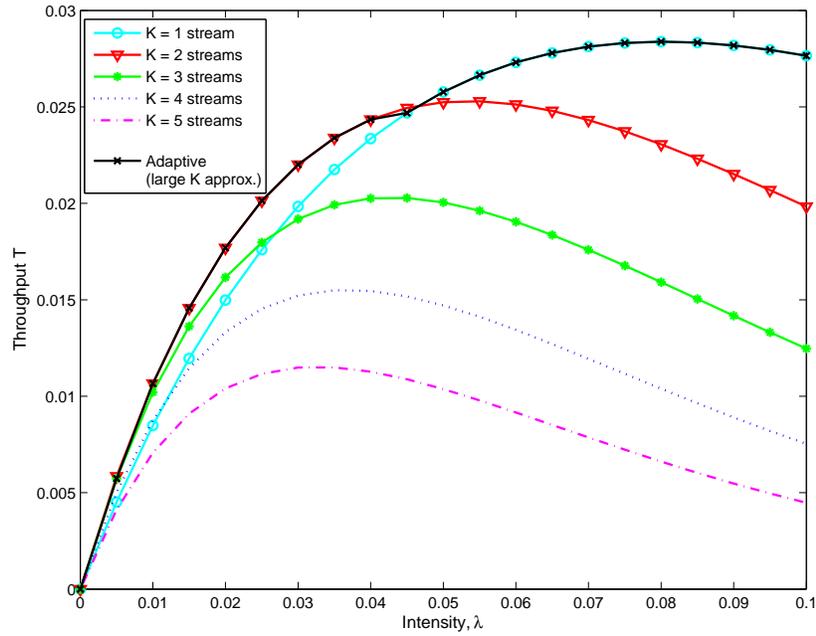}
\vspace{-5mm}
\caption{Network throughput vs. node density for $\alpha = 4$, $d = 1.5$, $M = 4$, $B = 10$ bits, $\beta = 1$ dB, and SNR = 15 dB.}
\label{fig10}
\end{figure}

In Fig. \ref{fig10}, we plot the network throughput as a function of the node density for different number of streams. As indicated by the above analysis, fully loaded SDMA ($K = M$) is detrimental for throughput performance, while multi-stream is superior only in sparse networks. An adaptive scheme that sets the number of streams based on the large $K$ approximation solution is also plotted. Interestingly, the adaptive scheme results in the optimal multi-stream transmission scheme within almost all range of intensity values $\lambda$. Furthermore, as expected, throughput increases when feedback quality is improved. We also plot the scheme in which the number of streams are adapted according to the large $K$ approximate value, which performs satisfactorily for low densities $\lambda$.

Regarding the transmission capacity, taking the derivative of (\ref{eq:TC}) with respect to $K$ and finding the stationary points, we have 
\begin{prop}
\textit{The optimal number of streams that maximizes the multi-stream transmission capacity is
\begin{equation*}
K^* = \min\left(\max\left(\left\lceil \nu \right\rceil,1\right),M\right) 
\end{equation*}
where $\nu$ is the solution of 
\begin{equation*}
K\frac{\partial \mathcal{I}_K}{\partial K}(KC_3 - C_4) + \mathcal{I}_K (C_4-2KC_3) = 0,
\end{equation*}
with $C_3 = \frac{\sigma^2\zeta_{\rm max}}{P} + \log(1+\beta\delta)$ and $C_4 = -\log(1-\epsilon) + \log(1+\beta\delta)$.}
\end{prop}
\begin{remark}
Using the large $K$ approximation, the optimal number of streams is given by
\begin{equation}
K^* = \left\lceil \frac{(1-2/\alpha)C_4}{(2-2/\alpha)C_3} \right\rceil.
\end{equation} 
\end{remark}

\begin{figure}[ht]
\centering
\includegraphics[width=5in]{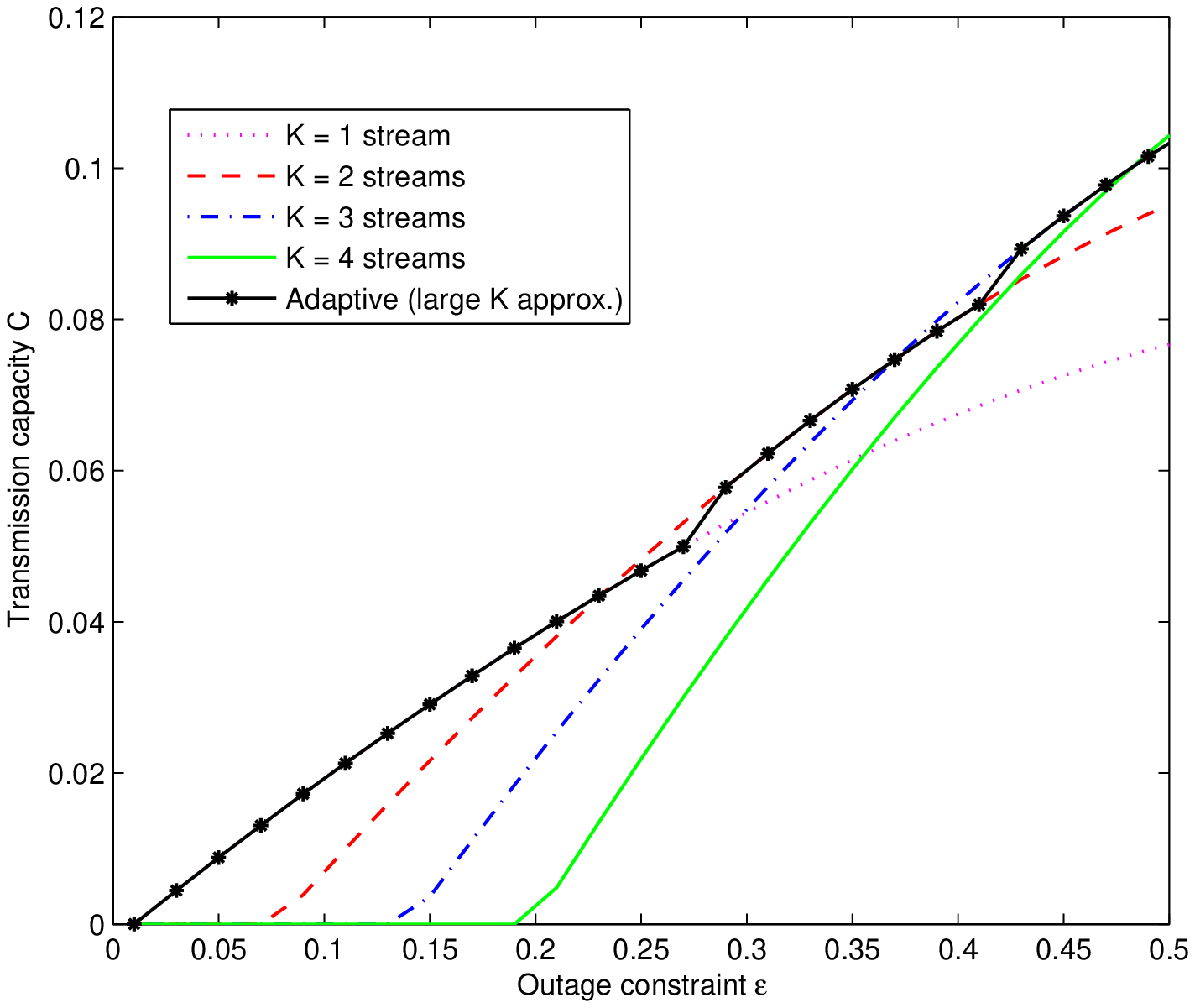}
\vspace{-5mm}
\caption{Multi-stream transmission capacity vs. outage constraint for $\alpha = 4.5$, $M = 4$, $B = 12$ bits, $\beta = 1$ dB, and SNR = 20 dB.}
\label{fig11}
\end{figure}

In Fig. \ref{fig11}, we plot the multi-stream transmission capacity vs. outage constraint for different number of streams employed. For most relevant parameters, single-stream transmission is optimal, even for a large number of bits. Similarly to the throughput case, the adaptive scheme based on the simple large $K$ approximate solution performs satisfactorily in a wide range of $\epsilon$ values.



\section{Conclusions}
\label{sec:conclusions}
We investigated the performance of zero-forcing precoding with limited feedback in single-hop ad hoc networks under a broad set of metrics and scenarios. The main takeaway of this paper is that SDMA may not be a wise use of transmit antennas in decentralized networks with both self and other user interference, as single-stream transmission maximizes both outage-based and average throughput in most practically relevant scenarios. In other words, a high density of single-stream communication links may be preferable than SDMA transmission with quantized channel state information.
Our analytical framework enables us to quantify the effect of the residual multiuser interference due to quantized CSIT on network-wide performance and to properly adjust the number of streams and the feedback bit scaling to achieve certain level of rate performance. A key finding that the per-user feedback load must be increased almost linearly with the number of antennas and logarithmically with the target SINR. The techniques developed here are also relevant for the analysis of partial CSIT-based linear precoding in emerging heterogeneous network paradigms, including femtocells, relays, picocells, and WiFi hotspots.

Further extensions to this work could include different limited feedback precoding schemes (e.g. MMSE) or how to exploit multiple receive antennas for interference cancelation. Future work could consider multi-hop networks with opportunistic routing and investigate potential SDMA gains into end-to-end performance (e.g. progress-rate-density). It would be also of interest to explore how opportunistic user selection affects the spatial reuse and rate, as well as how to properly design CSIT in networks with spatial randomness.



\appendix{

\subsection{Proof of Theorem~\ref{Th:Pout}}
\label{Th:Pout_proof}

Let $\mathcal{L}_p$ denote the Laplace transform of the normalized aggregate interference from the Poisson field of interferers (inter-cluster) $\overline{I}_p = \sum_{i \in \Phi(\lambda)} S_{ik} |X_i|^{-\alpha}$ with fading marks $S_{ik}$, defined as $\mathcal{L}_p(s) = \mathbb{E}_{\overline{I}_p}\left[e^{-s\overline{I}_p}\right] = \int_{0}^{\infty}e^{-sp}f_{\overline{I}_p}(p)dp$. The Laplace transform of the normalized intra-cluster interference $\overline{I}_q$ due to multiuser transmission with quantized CSIT is denoted by $\mathcal{L}_q(s)$. Define also the random variable $Y = \overline{I}_p + \overline{I}_q$.

The outage probability is given by $\pkout = 1 - \mathbb{P}\left\{\sinr_k \geq \beta_k \right\}$, which can be rewritten as
\begin{equation*}
\pkout  =  1 - \mathbb{P}\left\{\frac{\rho S_{0k} d_k^{-\alpha}}{I_p  + I_q + \sigma^2} \geq \beta_k \right\}   = 1 - \mathbb{P}\left\{S_{0k} \geq \beta_k d_k^{\alpha}(\overline{I}_p + \overline{I}_q + \sigma^2/\rho) \right\}. 
\end{equation*}
The channel gain is given by $S_{0k} = \left|\mathbf{h}_{k}\mathbf{w}_{k}\right|^2 = \left\|\mathbf{h}_{k}\right\|^2 \left|\overline{\mathbf{h}}_{k}\mathbf{w}_{k}\right|^2 = \left\|\mathbf{h}_{k}\right\|^2 \mathcal{B}(1,M-1)$, where $\mathcal{B}(1,M-1)$ is a Beta distributed random variable (r.v.) with shape parameters $(1, M -1)$ and independent of $\left\|\mathbf{h}_{k}\right\|^2$ \cite{Yoo_JSAC}. The term $\left\|\mathbf{h}_{k}\right\|^2$ is distributed as a chi-squared r.v. with $2M$ degrees of freedom denoted as $\chi_{2M}^{2}$.
Thus, $S_{0k} \sim \rm exp(1)$ (exponentially distributed with unit mean). Denoting $\zeta_k = \beta_k d_k^\alpha$, we have 
\begin{eqnarray}
\pkout  \stackrel{(a)} =  1 - \mathbb{E}\left[\exp(-\zeta_k(\overline{I}_p + \overline{I}_q + \sigma^2/\rho) ) \right] \stackrel{(b)} = 1 - \mathcal{L}_{p}(\zeta_k)\mathcal{L}_{q}(\zeta_k)e^{-\frac{\zeta_k\sigma^2}{\rho}}
\label{LpLq}
\end{eqnarray}
where step (a) is reached by conditioning on the aggregate interference $\overline{I}_p + \overline{I}_q$ and (b) by the independence of the interference terms.

The interferer marks in $\overline{I}_p$ are chi-squared distributed with degrees of freedom $S_{ik} = \left\|\mathbf{h}_{ik}\mathbf{W}_{i}\right\|^2 \sim \chi_{2K}^{2}$ since it is the sum of $K$ i.i.d. exponential random variables. Thus, the Laplace transform for a Poisson shot noise process in $\mathbb{R}^2$ with i.i.d. $\chi_{2K}^{2}$ distributed marks is given by \cite{Kin93}
\begin{eqnarray}
\mathcal{L}_p(s)  =  \mathbb{E}_{\Phi}\left[e^{-s\sum_{i \in \Phi(\lambda)} S_i |X_i|^{-\alpha})}\right]  =  \exp\left\{-\lambda\int_{\mathbb{R}^2}1 -\mathbb{E}_S\left[ e^{-s S |x|^{-\alpha}}\right]\textrm{d}x \right\} = e^{-\lambda s^{\frac{2}{\alpha}} \mathcal{I}_K}
\label{LaplaceP}
\end{eqnarray}
where \begin{equation}
\label{laplace_g}
\mathcal{I}_K = \frac{2\pi}{\alpha}\displaystyle \sum_{m=0}^{K-1}\binom{K}{m}B\left(m+\frac{2}{\alpha},K-m-\frac{2}{\alpha}\right) 
\end{equation}
with $B(a,b) = \int_{0}^{1}t^{a-1}(1-t)^{b-1}dt = \frac{\Gamma(a)\Gamma(b)}{\Gamma(a+b)}$ being the Beta function.

For the marks in the interference term we have that $\left|\mathbf{h}_{k}\mathbf{w}_{j}\right|^2 = \left\|\mathbf{h}_{k}\right\|^2 \left|\overline{\mathbf{h}}_{k}\mathbf{w}_{j}\right|^2$. 
Denoting $\phi_k$ the angle between $\mathbf{h}_{k}$ and $\hat{\mathbf{h}}_{k}$, and decomposing the normalized channel vector as $\overline{\mathbf{h}}_{k} = (\cos\phi_k)\hat{\mathbf{h}}_{k} + (\sin\phi_k)\mathbf{v}_{k}$, we have that $\left|\mathbf{h}_{k}\mathbf{w}_{j}\right|^2 = \left\|\mathbf{h}_{k}\right\|^2 \left|\mathbf{v}_{k}\mathbf{w}_{j}\right|^2 \sin^2\phi_k$. Since $\mathbf{w}_{j}$ is isotropic within the hyperplane and independent of $\mathbf{v}_{k}$, the quantity $\left|\mathbf{v}_{k}\mathbf{w}_{j}\right|^2$ is beta $(1, M-2)$ distributed, i.e. $\left|\mathbf{v}_{k}\mathbf{w}_{j}\right|^2 \sim \mathcal{B}(1,M-2)$ \cite{Jindal_finiteFB} and independent of the quantization error $\sin\phi_k$. Thus, the normalized intra-cluster interference can be rewritten as
\begin{eqnarray} 
\overline{I}_q = d_k^{-\alpha}\left\|\mathbf{h}_{k}\right\|^2 \sin^2\phi_k\displaystyle \sum_{j \in \mathcal{K}, j \neq k} \mathcal{B}(1,M-2),
\end{eqnarray} 
where $\left\|\mathbf{h}_{k}\right\|^2$, $\phi_k$, and $\mathcal{B}(1,M-2)$ are all independent. For the quantization cell approximation, the term $X = \left\|\mathbf{h}_{k}\right\|^2 \sin^2\phi_k$ is a gamma distributed r.v. with shape $M-1$ and scale $\delta$, i.e. $X \sim {\rm Gamma}(M-1, \delta)$. Therefore $Y = X\mathcal{B}(1,M-2)$ is exponentially distributed with rate $1/\delta$ \cite{JunTCOM}. As $\overline{I}_q = d_k^{-\alpha}Z$, where $Z$ is the sum of $(K-1)$ i.i.d exponentially distributed r.v., i.e. $Z \sim {\rm Gamma}(K-1,\delta)$, the normalized aggregate interference becomes $\overline{I}_q \sim {\rm Gamma}(K-1,\delta d_k^{-\alpha})$. The Laplace transform of $\overline{I}_q$ is given by 
\begin{eqnarray}
\mathcal{L}_q(s) = \frac{1}{(1+sd_k^{-\alpha}\delta)^{K-1}}.
\label{LaplaceQ}
\end{eqnarray}
Substituting (\ref{LaplaceP}) and (\ref{LaplaceQ}) in (\ref{LpLq}) we obtain the result.

\subsection{Proof of Theorem~\ref{Th_Erg}}
\label{Th_Erg_proof}
For a random variable $X$ with probability density function $f_X(x)$ and cumulative distribution function (cdf) $F_X(x)$, we have
\begin{eqnarray*}
\mathbb{E}\{\log(1+X)\}  = \int_{0}^{\infty}\log(1+x)f_X(x)\textrm{d}x = \int_{0}^{\infty}\log(1+x)\textrm{d}\left[1 - F_X(x)\right]  \stackrel{(a)} =  \int_{0}^{\infty}\frac{1-F_X(x)}{1+x}\textrm{d}x
\end{eqnarray*}
where step (a) follows from integration by parts. The result is obtained by
\begin{equation*}
\mathcal{R}_k  = \int_{0}^{\infty}\frac{1-\mathcal{F}_{(x,\alpha)}^{(k)}}{1+x}\textrm{d}x = \int_{0}^{\infty}\frac{e^{-\frac{\sigma^2d_k^{\alpha} x}{\rho}}e^{-\lambda\mathcal{I}_{M}d_k^{2}x}}{(1+x)(1+\delta x)^{M-1}}\textrm{d}x, \ \ \ \text{with} \ C_1 = \frac{\sigma^2d_k^{\alpha} x}{\rho} \ \ \text{and} \ C_2 = \lambda\mathcal{I}_{M}d_k^{2}.
\end{equation*}

\subsection{Proof of Theorem~\ref{Th_ErgJ}}
\label{Th_ErgJ_proof}
First, we consider the following upper bounds to the average rate achieved by the $k$-th user:
\begin{eqnarray}
\mathcal{R}_k  & = & \mathbb{E}\left\{\log\left(1 + \frac{\rho \left|\mathbf{h}_{k}\mathbf{w}_{k}\right|^2 d_k^{-\alpha}}{\overline{I}_p  + \overline{I}_q + \sigma^2/\rho}\right)\right\} 
\stackrel{(a)} \leq \mathbb{E}\left\{\log\left(1 + \frac{\left|\mathbf{h}_{k}\mathbf{w}_{k}\right|^2d_k^{-\alpha}}{\overline{I}_p^{l}  + \overline{I}_q + \sigma^2/\rho}\right)\right\} \nonumber \\
& = & \mathbb{E}\left\{\log\left(\overline{I}_p^{l}  + \overline{I}_q + \frac{\sigma^2}{\rho} + \left|\mathbf{h}_{k}\mathbf{w}_{k}\right|^2d_k^{-\alpha}\right)\right\} - 
\mathbb{E}\left\{\log\left(\overline{I}_p^{l}  + \overline{I}_q + \frac{\sigma^2}{\rho} \right)\right\} \nonumber \\
& \stackrel{(b)} \leq & \log\left(\mathbb{E}\left\{\overline{I}_p^{l}  + \overline{I}_q + \frac{\sigma^2}{\rho} + \left|\mathbf{h}_{k}\mathbf{w}_{k}\right|^2d_k^{-\alpha}\right\} \right) - \mathbb{E}\left\{\log\left(\overline{I}_q \right)\right\} \nonumber  \\
& \stackrel{(c)} \leq & \log\left(\mathbb{E}\left\{\overline{I}_p^{l}\right\}  + \mathbb{E}\left\{\overline{I}_q\right\} + \frac{\sigma^2}{\rho} + d_k^{-\alpha}\mathbb{E}\left\{\left|\mathbf{h}_{k}\mathbf{w}_{k}\right|^2\right\} \right) - \mathbb{E}\left\{\log\left(d_k^{-\alpha} \left|\mathbf{h}_{k}\mathbf{w}_{j}\right|^2 \right)\right\} 
\label{ub:rate}
\end{eqnarray}
where $\overline{I}_p = I_p/\rho$ and $\overline{I}_q = I_q/\rho$ are given in (\ref{eq:Ip}) and (\ref{eq:Iq}), respectively.
In (a) we consider a lower bound to the inter-cluster interference, denoted as $\overline{I}_p^{l}$, and in (b) we apply Jensen's inequality to the minuend and neglect the noise and inter-cluster interference terms in the subtrahend. Step (c) follows from the independence of the interference terms and the received signal and by considering only one of the intra-cluster interference terms.

The normalized interference $\overline{I}_q \sim {\rm Gamma}(K-1,\delta d_k^{-\alpha})$, thus we have $\mathbb{E}\left\{\overline{I}_q\right\} = d_k^{-\alpha}\delta (M-1)$. The channel gain  $\left|\mathbf{h}_{k}\mathbf{w}_{k}\right|^2$ is exponentially distributed with mean one, i.e. $\mathbb{E}\left\{\left|\mathbf{h}_{k}\mathbf{w}_{k}\right|^2\right\} = 1$. 
In order to derive a lower bound, we neglect the contribution of the closest interferer to the inter-cluster interference. Using techniques from \cite{PZF}, the lower bound is given by $\overline{I}_p^{l} = \left(\pi\lambda d_k^{2}\right)^{\frac{\alpha}{2}} \left(\left\lceil \frac{\alpha}{2}\right\rceil-1\right)^{-\frac{\alpha}{2}}$.

For the second term in (\ref{ub:rate}), we need to compute
\begin{equation}
\mathbb{E}\left\{\log\left(\left|\mathbf{h}_{k}\mathbf{w}_{j}\right|^2 \right)\right\} = \mathbb{E}\left\{\log\left(\left\|\mathbf{h}_{k}\right\|^2\left|\overline{\mathbf{h}}_{k}\mathbf{w}_{j}\right|^2 \right)\right\} = \mathbb{E}\left\{\log\left(\left\|\mathbf{h}_{k}\right\|^2 \right)\right\} + \mathbb{E}\left\{\log\left(\left|\overline{\mathbf{h}}_{k}\mathbf{w}_{j}\right|^2 \right)\right\}.\\
\end{equation}
For the channel norm we have $\left\|\mathbf{h}_{k}\right\|^2 \sim \chi_{2M}^2$, thus $\mathbb{E}\left\{\log\left(\left\|\mathbf{h}_{k}\right\|^2 \right)\right\} = \int_0^{+\infty}\log x\cdot \frac{x^{M-1}e^{-x}}{\Gamma(M)} \textrm{d}x = \psi(M)$.
Then, $\mathbb{E}\left\{\log\left(\left|\overline{\mathbf{h}}_{k}\mathbf{w}_{j}\right|^2 \right)\right\} = \mathbb{E}\left\{\log\left(\sin^2\phi_k\right)\right\} + \mathbb{E}\left\{\log\left(\mathcal{B}(1,M-2)\right)\right\}$, with
\begin{eqnarray*}
\mathbb{E}\left\{\log\left(\sin^2\phi_k\right)\right\} = \int_0^{-\infty}\log(x) {\textrm d}F_{\sin^2\phi_k}(x) = 2^{B}(M-1)\int_0^{\delta}\log(x) x^{M-2}\textrm{d}x = - \frac{1+B\log2}{M-1}
\end{eqnarray*}
and 
\begin{eqnarray*}
\mathbb{E}\left\{\log\left(\mathcal{B}(1,M-2)\right)\right\} = \int_0^{1}\log x\cdot \frac{(1-x)^{M-3}}{B(1,M-2)} \textrm{d}x = - H_{M-2}
\end{eqnarray*}
where the cdf of the quantization error $\sin^2\phi_k$ is given in \cite{Yoo_JSAC}.
By substituting the above quantities to (\ref{ub:rate}) and after some manipulations, we obtain (\ref{eq:th4}).

\subsection{Proof of Lemma~\ref{Holder_Erg}}
\label{Holder_Erg_proof}
The proof of this lemma is based on H\"{o}lder's inequality \cite{HardyIneq}, which states that for two measurable functions $f$ and $g$ defined on a Hilbert space $\textsl{S}$ and $1 < p, q < \infty$ with $1/p + 1/q = 1$
\begin{equation} 
\label{HolderInq_eq}
\int_{\textsl{S}} \left|f(x)g(x)\right|\textrm{d}x = \left(\int_{\textsl{S}} \left|f(x)\right|^p\textrm{d}x\right)^{1/p} \left(\int_{\textsl{S}} \left|g(x)\right|^q\textrm{d}x\right)^{1/q}.
\end{equation}

The result is obtained by applying twice (\ref{HolderInq_eq}) for the following decreasing real-valued and bounded functions: first $f(x) = e^{-C_1x}e^{-C_2x^{\frac{2}{\alpha}}}$ and $g(x) = \frac{1}{(1+x)(1+\delta x)^{M-1}}$ and then for $f(x) = e^{-C_1x}$ and $g(x) = e^{-C_2x^{\frac{2}{\alpha}}}$. At each step, we optimize $p, q$ in order to obtain the tightest possible upper bound. Using the following forms
\begin{eqnarray*}
\mathcal{I}_1 & = & \int_{0}^{\infty}e^{-ax}\textrm{d}x = \frac{1}{a}, \ \ \ \ \ \ \ \ \ \ \  \mathcal{I}_2  =  \int_{0}^{\infty}e^{-bx^{2/\alpha}}\textrm{d}x = b^{-\alpha/2} \Gamma(1 + \alpha/2),\\
\mathcal{I}_3 & = & \int_{0}^{\infty}\frac{\textrm{d}x}{(1+x)^c(1+\delta x)^{g}} = \frac{\pi \csc(g\pi)(\delta-1)^{1-c-g}\delta^{c}B(c,g-1) + _2F_1(1,c,2-g,\frac{1}{\delta})}{\delta g},
\end{eqnarray*}
where $\csc(\cdot)$ denotes the cosecant and $_2F_1(x,y,z,w)$ the Gauss hypergeometric function, and after some algebraic manipulations, we obtain (\ref{holder_I}).    


}

\bibliographystyle{IEEEtran}
\bibliography{adhoclit}

%

\vspace{-30mm}

\centering

\end{document}